\documentclass[conference]{IEEEtran}
\IEEEoverridecommandlockouts
\usepackage{cite}
\usepackage{amsmath,amssymb,amsfonts}
\usepackage[colorlinks,urlcolor=blue,linkcolor=blue,citecolor=blue]{hyperref}
\usepackage{graphicx}
\usepackage{textcomp}
\usepackage{xcolor}
\usepackage[numbers,sort&compress]{natbib}
\usepackage{color,array}
\usepackage{caption}
\usepackage{mathtools} 
\usepackage{soul}
\usepackage{fancyhdr}
\usepackage{multirow}
\usepackage{multicol}
\usepackage{enumitem}
\usepackage[noabbrev]{cleveref}
\usepackage{algorithm}
\usepackage{algpseudocode}

\algnewcommand{\Inputs}[1]{%
  \State \textbf{Inputs:}
  \Statex \hspace*{\algorithmicindent}\parbox[t]{.8\linewidth}{\raggedright #1}
}
\algnewcommand{\Initialize}[1]{%
  \State \textbf{Initialize:}
  \Statex \hspace*{\algorithmicindent}\parbox[t]{.8\linewidth}{\raggedright #1}
}
\algnewcommand{\Data}[1]{%
  \State \textbf{Data:}
  \Statex \hspace*{\algorithmicindent}\parbox[t]{.8\linewidth}{\raggedright #1}
}

\fancypagestyle{github}{\fancyhf{} \fancyfoot[L]{*\url{https://github.com/mahindrautela/BOPINN}}}

\begin{document}

\title{Bayesian optimized physics-informed neural network for estimating wave propagation velocities
\thanks{*Corresponding authors}
\thanks{J.S. acknowledge funding from the Accelerated Materials Development for Manufacturing Program at A*STAR via the AME Programmatic Fund by the Agency for Science, Technology and Research under Grant No. A1898b0043.}
}

\author{\IEEEauthorblockN{Mahindra Rautela*}
\IEEEauthorblockA{\textit{Los Alamos National Laboratory}\\
Los Alamos, US \\
mahindrautela@gmail.com}
\and
\IEEEauthorblockN{S. Gopalakrishnan}
\IEEEauthorblockA{\textit{Dept. of Aerospace Engineering} \\
\textit{Indian Institute of Science}\\
Bangalore, India \\
krishnan@iisc.ac.in}
\and
\IEEEauthorblockN{J. Senthilnath*}
\IEEEauthorblockA{\textit{Institute for Infocomm Research} \\
\textit{Agency for Science, Technology and Research}\\
Singapore \\
J\_Senthilnath@i2r.a-star.edu.sg}
}
\maketitle

\begin{abstract}
In this paper, we propose a novel inverse parameter estimation approach called Bayesian optimized physics-informed neural network (BOPINN). In this study, a PINN solves the partial differential equation (PDE), whereas Bayesian optimization (BO) estimates its parameter. The proposed BOPINN estimates wave velocity associated with wave propagation PDE using a single snapshot observation. An objective function for BO is defined as the mean squared error (MSE) between the surrogate displacement field and snapshot observation. The inverse estimation capability of the proposed approach is tested in three different isotropic media with different wave velocities. From the obtained results, we have observed that BOPINN can accurately estimate wave velocities with lower MSE, even in the presence of noisy conditions. The proposed algorithm shows robust predictions in limited iterations across different runs. 
\end{abstract}


\section{Introduction}
The study of wave propagation finds applications in different fields of science and engineering. Wave behavior in a media is governed by the wave equation, which provides the spatiotemporal evolution of a field. Numerical schemes like finite element methods (FEM) are popularly used to solve the wave equations in complex media. One of the significant challenges faced by FEM is their frequency-dependent meshing, which demands dense mesh for high-frequency wave propagation \cite{rautela2023deep}. This consequently requires higher computational resources and time. The inverse problem of estimating wave speed from measured observations becomes complicated if the forward problem faces the aforementioned challenges. In addition, the inverse problem has its issues in finding a unique solution. 
\begin{figure*}[ht!]
\centering
\includegraphics[width=0.75\textwidth]{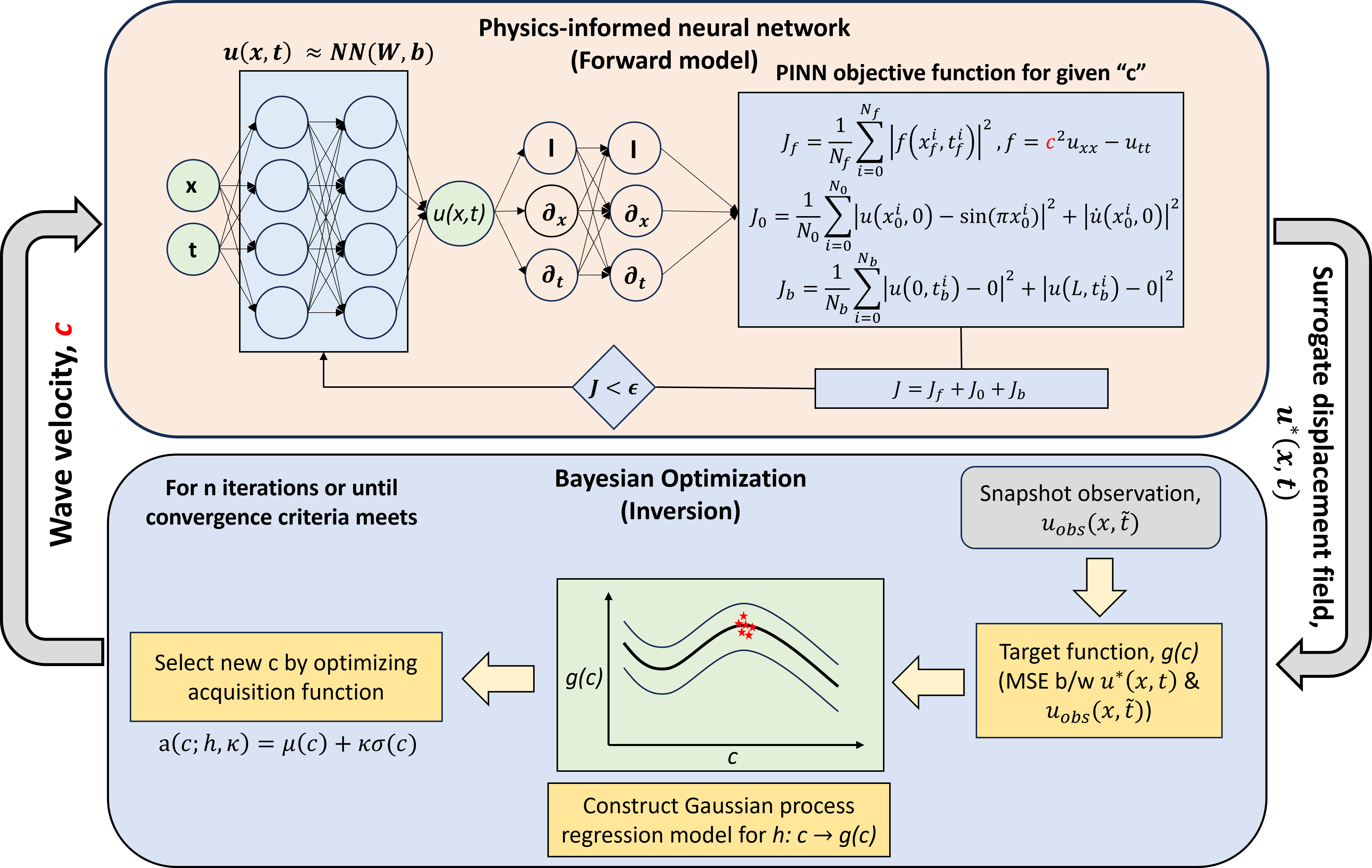}
\caption{The flow diagram of the BOPINN for inverse estimation of wave velocity}
\label{fig:bopinnmethod}
\end{figure*}

In recent years, machine learning (ML) and deep learning (DL) has emerged as a promising solver for inverse problems using guided wave propagation \cite{rautela2021inverse}. The forward problem is solved using a fast numerical solver like spectral FEM \cite{rautela2020ultrasonicIEEE}, semi-analytical FEM \cite{wang2022real}, stiffness matrix method \cite{rautela2021inverse}. The data collected is used to train ML/DL models to solve the inverse problem. The bottleneck becomes the data collection and storage process (high memory utilization). With the introduction of physics-informed neural networks (PINNs), some of the challenges, like frequency-dependent meshing and data collection processes, can be circumvented. A PINN presents a dense neural network (DNN) based surrogate model with a surrogate field to solve the partial differential equations (PDEs) \cite{raissi2019physics}. 

PINNs are utilized to solve the wave equation in the literature. The frequency-domain scattered pressure wavefields are predicted for a transversely isotropic Earth's layered structure with a vertical axis of symmetry \cite{song2021solving}. In another work, a two-dimensional acoustic wave equation is solved for homogeneous, layered, and Earth-realistic spatially varying velocity models \cite{moseley2020solving}. The above-mentioned works show that PINNs can learn wave propagation behavior in different media. The inverse problem of estimation of wave velocity is studied in Ref.~\cite{shukla2020physics,karimpouli2020physics,rasht2022physics,rathod2022physics}. For this, another neural network is defined along with observed data as input to the entire PINN architecture \cite{rasht2022physics}. Apart from the additional complexity of the network and higher training time to estimate the inverse parameters, the network needs to be retrained for new observations when deployed in online settings. Although the PINN model is deterministic, a demerit exists in the form of uncertainty quantification while estimating the inverse parameters. The uncertainties are incorporated through different approaches in the PINNs. A probabilistic latent variable model (variational autoencoders) is used as a surrogate instead of a deep neural network (DNN) and is trained via adversarial inference procedure \cite{yang2019adversarial}. A combination of DNN and arbitrary polynomial chaos is used to perform uncertainty quantification for both forward and inverse problems \cite{zhang2019quantifying}. Bayesian neural network (BNN) is utilized as the surrogate model to introduce uncertainty for both forward and inverse solutions \cite{yang2021b}. However, BNN comes with high computational costs \cite{zhang2019quantifying}. In Ref.~\cite{zhang2019quantifying,yang2021b}, dropouts are also incorporated to quantify the uncertainty in the DNN. They act as regularizers and help in avoiding overfitting. They are the computationally inexpensive way to include uncertainty in the networks. Also, the dropouts in DNNs are considered to be approximating a probabilistic deep Gaussian process \cite{gal2016dropout}. 

In the literature, inverse parameter estimation is also performed using optimization or search scheme in a model-calibration setting \cite{rautela2023real}. In this, an objective function with an unknown parameter is formulated using an error function between the numerical solution of the forward model and experimental observations. Gradient-based and gradient-free methods are popularly used to solve for unknown model parameters. However, the gradient-based method gets trapped in local optima and cannot be used when the objective function lacks a function form. Commonly used gradient-free, global optimization methods to address inverse problems are evolutionary optimization (EO) and Bayesian optimization (BO). The EO methods are population-based metaheuristic search techniques that require large amounts of data to converge to global optima, which makes them computationally expensive  \cite{dutta2022extracting}. On the other hand, Bayesian optimization is a probabilistic approach with limited data and dimensions, which can converge to global optima  \cite{krishnamoorthy2023model}. Also, BO brings uncertainty quantification in estimating the parameters \cite{zhang2022uncertainty}.

In this paper, we propose a novel Bayesian-optimized physics-informed neural network (BOPINN) to estimate wave velocity in isotropic materials. BOPINN consists of a forward PINN-based PDE solver to calculate the surrogate displacement field of wave propagation and a BO-based inversion scheme to estimate wave propagation velocity. We use a single noisy snapshot observation (simulated) as a representative experimental measurement. The Bayesian optimizer runs over the black-box PINN model and sequentially queries it adaptively at different points within the bounds of the domain until it converges to the true wave velocity \cite{raina2023robotic}. In order to incorporate uncertainties, we add dropouts in the forward PINN model, and the Gaussian process regression model in BO introduces uncertainties in the inverse estimation. The detailed methodology of BOPINN is presented in Fig.~\ref{fig:bopinnmethod}. BOPINN presents a decoupled approach, where the forward and inverse problems are separately formulated, contrary to the coupled approach where both models are stitched together in a common architecture \cite{rasht2022physics}. With this study, we bring an alternative approach to solve inverse problems. We consider BOPINN to be a more general formulation that can be used for estimating multiple parameters (a vector) and fields in different types of PDEs. However, in this study, we validate the proof of the concept using a single PDE with a single unknown parameter. 

We test the capability of the proposed technique for three test cases, i.e., three different isotropic media with different wave velocities. The robustness of the proposed method is evaluated through different repeated runs, which provides uncertainty estimation for different test cases. BOPINN offers many advantages, like simplicity in the forward model architecture, as it involves a single neural network when compared to the widely used dual neural network method in the literature \cite{rasht2022physics}. It is a probabilistic inversion scheme that additionally provides the uncertainty in estimation \cite{raina2023rusopt}. The methodology can be adapted in online inversion settings as it does not involve retraining with new experimental observations. 

We present a new paradigm to solve inverse problems by bringing an amalgamation of PINNs and BO. The main contributions of this paper are:
\begin{itemize}
    \item Developing a more general probabilistic methodology to estimate the parameter of PDEs with uncertainty quantification.
    \item Solving forward and inverse problems in wave propagation using PINN and BO, respectively.
    \item Using a single noisy snapshot observation for wave velocity estimation in different isotropic media.
    \item Presenting a computationally inexpensive method that can provide accurate and robust estimates within a limited number of iterations.
\end{itemize}

The paper is organized as follows: Section-\ref{sec:methodolgy} consists of the definition and formulation of PINN, BO, and BOPINN. Section-\ref{sec:results} contains different results following with discussions. The paper is concluded in Section-\ref{sec:conclusion}.

\section{Methodology} \label{sec:methodolgy}
The proposed BOPINN is used to estimate the parameters of a PDE, particularly the velocity of wave propagation. As highlighted in Fig.~\ref{fig:bopinnmethod}, BOPINN integrates PINN (forward model) and BO (inverse model). The BO sequentially queries the black-box PINN model in such a way that the unknown wave velocity approaches true velocity.

\subsection{Physics-informed neural networks}
The wave equation is a linear second-order hyperbolic PDE that describes wave motion in a medium. Wave equation finds applications in continuum mechanics, geophysics, electromagnetism, quantum mechanics, plasma physics, general relativity, and many other scientific and engineering disciplines \cite{achenbach2012wave}. Mathematically, the wave equation in a homogeneous and isotropic medium given by, 
\begin{equation}\label{eq:wave3d}
    \nabla^2 u = \frac{1}{c^2} \frac{\partial^2 u}{\partial t^2} \\
\end{equation}
where $u(s,t)$ is the displacement field as a function of space $s = (x,y,z)$ and time $t$, and $c$ is the wave speed in the medium. The one-dimensional wave equation is a particular case of Eq.~\ref{eq:wave3d} describing a unidirectional wave motion, $u(x,t)$ and can be written as
 \begin{equation}\label{eq:wave1d}
     \frac{\partial^2 u}{\partial x^2} = \frac{1}{c^2} \frac{\partial^2 u}{\partial t^2}
 \end{equation}

The above equation needs two initial and two boundary conditions for its solution. We define the Dirichlet type boundary condition where the field variable is defined at the boundaries \cite{rautela2021simulation}. The domain, initial and boundary conditions for the Eq.~\ref{eq:wave1d} can be written as
\begin{align}
	x \in [0,L], & \hspace{2mm} t \in [0,T] \label{eq:domain}\\
	u(x,t = 0) = -\sin(\pi x), & \hspace{2mm} \dot{u}(x,t=0) = 0 \label{eq:ic} \\
	u(x=0, t) = 0 , & \hspace{2mm} u(x = L, t) = 0 \label{eq:bc}
\end{align}

The displacement field is zero at both the boundaries (x=-L and x=+L). We have provided the initial condition for displacement as a sinusoidal spatial function, and the initial condition for the velocity field is zero. Eq.~\ref{eq:wave1d} can be written in the form of a PINN as $f(x,t):=c^2 u_{xx}-\ddot{u} = 0$. A dense neural network is used as a surrogate displacement field in terms of the domain to approximate the true displacement field, i.e., $u(x,t) \thickapprox NN(W,b)$ such that it satisfies $f$, domain, initial and boundary conditions. In order to achieve it, a loss function is defined, which is a summation of all the mean squared errors (MSE) on $f$ ($J_f$), initial ($J_0$), and boundary conditions ($J_b$).
\begin{gather}
	J_{f} =\frac{1}{N_f} \sum_{i=0}^{N_f}|f(x^i_f,t^i_f)|^2 \\
	J_{0} =\frac{1}{N_0} \sum_{i=0}^{N_0}\bigg(|u(x^i_0,0) - \sin(\pi x^i_0)|^2 + |\dot{u}(x^i_0,0)|^2 \bigg) \\
	J_{b} =\frac{1}{N_b} \sum_{i=0}^{N_b}\bigg(|u(0,t^i_b)|^2 + |u(L,t^i_b)|^2\bigg)\\
        J = J_{f} + J_{0} + J_{b}
\end{gather}
where, $\{x^i_0,t^i_0\}^{N_0}_{i=1}$ is the initial data, $\{x^i_f,t^i_f\}^{N_f}_{i=1}$ and $\{x^i_b,t^i_b\}^{N_b}_{i=1}$ are the collocation points for $f(x,t)$ and boundaries. The optimization problem can be written as
\begin{equation} \label{eq:PINNopt}
    u^*(x,t) = \arg \min_{u(x,t)} J(u(x,t),c)
\end{equation}

A neural network is trained to minimize the loss function $J(u(x,t),c)$ using mini-batch gradient descent with back-propagation. During this process, the optimal parameters of the neural network i.e., weights (W) and biases (b) are learned. The optima of the objective function gives surrogate displacement field $u^*(x,t) \thickapprox NN(W^*,b^*)$, which is necessarily the solution of the PDE \cite{karniadakis2021physics}.

\subsection{Bayesian Optimization}
The solution of the wave equation depends on the wave velocity parameter ($c$), which relies on the properties of the medium. The PINN-based methodology provides the forward solution, i.e., displacement field $u(x,t)$ considering $c$ as a constant value. However, the goal of the inverse problem is to estimate $c$ provided some measurement on the field or its derivatives using BO. In this work, we have used a single snapshot of the wave propagation (axial displacement) at different locations at a particular time (at $t=\tilde{t}$). We have simulated this observation using an analytical expression but added white noise to it. The data can be assumed to be taken at different locations by an array of sensors, similar to seismology \cite{rasht2022physics}. The inverse problem of estimating wave velocity can be thought of as a model-calibration problem where the unknown parameter (wave velocity $c$) of the model can be estimated using some observation \cite{rautela2023real}. In this study, the inversion scheme is required to perform
\begin{gather} 
    c^* = \arg \min_{c \in C} g(c) = \arg \max_{c \in C} -g(c)  \\
    c^* = \arg \max_{c \in C} -\parallel\arg\min_{u(x,t)} J(u(x,t),c) |_{t=\tilde{t}} - u_{obs}(x,\tilde{t})\parallel_2 \label{eq:BO}
\end{gather}
where $g(c)$ is the target function and $c^*$ is the optimum value. Here, $c^*$ is the estimated wave velocity and $g(c)$ is the mean squared loss function between displacement fields coming from the PINN wave model and snapshot observation at $t=\tilde{t}$.

\begin{algorithm}[h!]
\caption{BOPINN algorithm}
\begin{algorithmic}
    \Inputs{n, N, C, parameters of PINN, AF}
    \Data{$u_{obs}(x,\tilde{t})$}
    \Initialize{
    $c = \{c_1,c_2,...,c_n\} \subset C$ \\
    PINN: $u^* \gets \arg \min_{u} J(u,c)$\\
    $y = g(c) \gets -(u^*(x,t)|_{t=\tilde{t}} - u_{obs}(x,\tilde{t}))^2$ \\
    Train GPR $h:c \mapsto g(c)$ 
    }
    \While{$i \leq N$}
    \State AF: $c' \gets \arg \max_{c \in C} a(c;h,\kappa)$ 
    \State $y' \gets g(c')$
    \State $c \gets c \cup \{c'\}$
    \State $y \gets y \cup \{y'\}$
    \State Retrain GPR $h:c \mapsto g(c)$    
    \EndWhile   
\end{algorithmic}
\label{alg:algo}
\end{algorithm}

\begin{figure*}[ht!]
\centering
\includegraphics[width=0.6\textwidth]{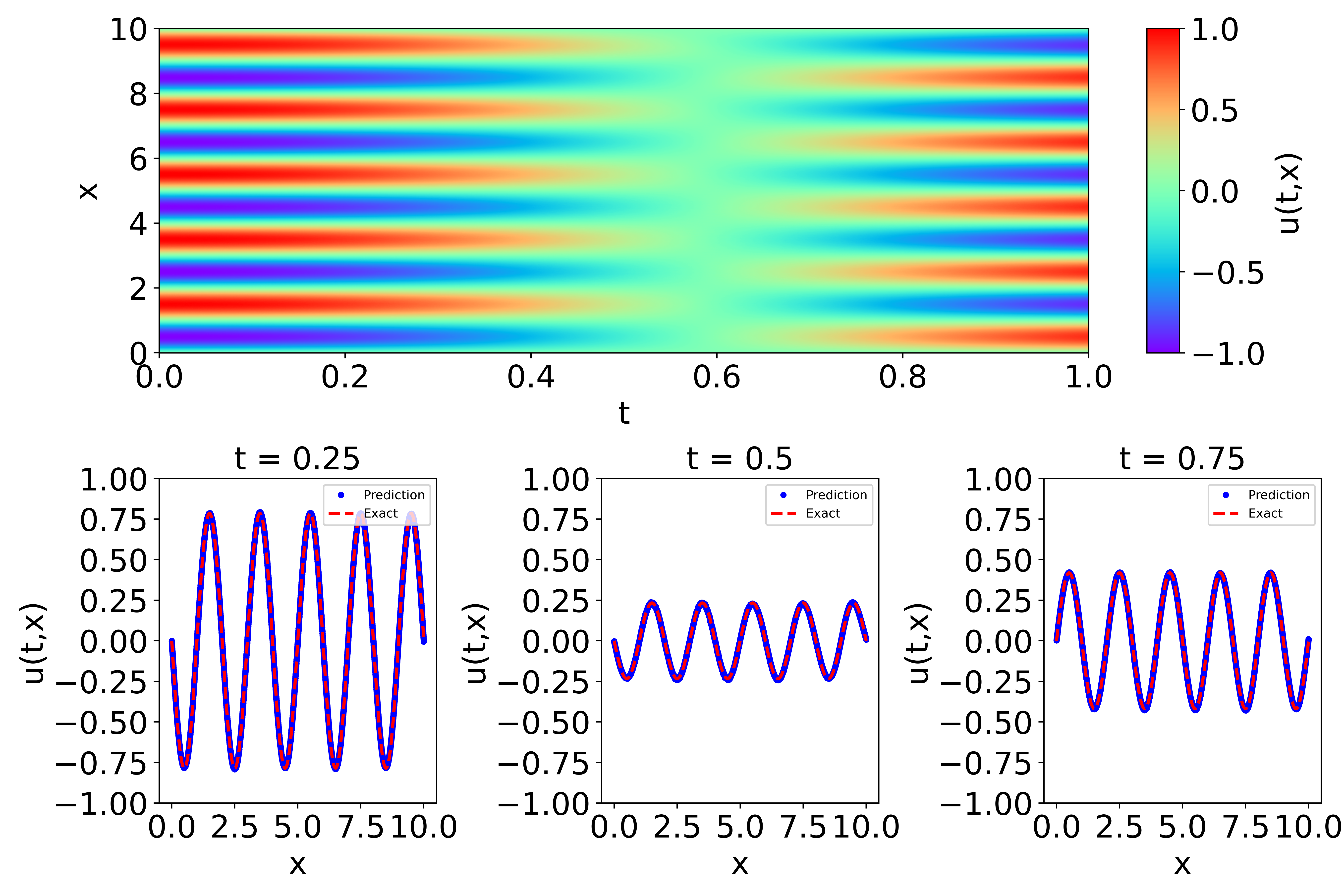}
\caption{Solution of PDE at $c=8500$ m/s. The top plot shows normalized spatiotemporal displacement field and the bottom three plots represents displacement field at different time slices.}
\label{fig:pinnsol}
\end{figure*}

One important thing to note here is that we have a black-box PINN-based forward model, and calibrating it with gradient-based optimization schemes is infeasible \cite{rautela2023real}. On the other hand, BO is a more appealing candidate for the problem at hand as BO is a gradient-free global optimization scheme that can be adapted in an online setting to solve the inverse problem \cite{sha2020applying}. The entire process of BOPINN is enumerated with Algorithm~\ref{alg:algo}. The objective function of BO is called the target function ($g(c)$), which is defined as the mean squared loss between the surrogate displacement field and snapshot observation at a particular time. BO has two main components, i.e., a Bayesian statistical model, i.e., Gaussian process regression (GPR), and an acquisition function (AF) \cite{snoek2012practical}. The GPR (defined by $h$) models the target function ($g(c)$) and provides a Bayesian prior probability distribution using initial randomly chosen points for $c$. An acquisition function decides where to sample the next point $c$ by calculating the maxima of the function. The posterior distribution, i.e., GPR, is retrained and updated every iteration. All the old and new $c$ and $g(c)$ are also concatenated. This interplay between the Bayesian model and acquisition function runs until convergence or a given number of iterations. Due to the probabilistic framework of BO, the algorithm can be run multiple times to give uncertainty bounds in estimating the parameters.

We have used Upper confidence bound (UCB) as the acquisition function, defined mathematically as
\begin{equation}\label{eq:acqfn}
    a(c;h,\kappa) = \mu(c) + \kappa \sigma(c)
\end{equation}
where, $\mu(c)$ and $\sigma(c)$ are the mean and standard deviation of the GPR ($h$). $\kappa$ defines the trade-off between exploration and exploitation \cite{snoek2012practical}. Higher value of $\kappa$ scales the standard deviation that increase the uncertainty and rewards the exploration in uncharted search space. A lower $\kappa$ selects better performing solutions and enables more exploitation. The $\kappa$ can also be decayed dynamically during the iterations, which provides more exploration during the initial iterations and more exploitation towards the latter iterations.

\begin{table*}
\small
\centering
\captionsetup{justification=centering}
\setlength{\belowcaptionskip}{0pt}
\caption{Target function-velocity results of BOPINN for all the three cases. The columns shows best optimal, mean optimal and the standard deviation across 10 different runs.}
\label{tab:BOPINNtab}
\addtolength{\tabcolsep}{-1pt}
\begin{tabular}{|c|c|c|c|} 
    \hline
    & True c & Best Optimal (TF*, c*) & Mean Optimal and Standard Deviation (TF*, c*) \\
    \hline
    Case-1 & \textbf{0.2000} &(-6.565e-05, \textcolor{red}{0.1969}) & (-7.280e-05 $\pm$ 4.748e-06, 0.2016 $\pm$ 0.0163) \\ 
    \hline
    Case-2 & \textbf{0.5500} &(-8.553e-05, \textcolor{red}{0.5512}) & (-0.00038 $\pm$ 0.000498, 0.5190 $\pm$ 0.0788) \\
    \hline
    Case-3 & \textbf{0.8500} &(-6.238e-05, \textcolor{red}{0.8495}) & (-7.076e-05 $\pm$ 7.369e-06, 0.8480 $\pm$ 0.0073) \\ 
    \hline
\end{tabular}
\end{table*}

\begin{figure*}
    \centering
    \begin{minipage}[b]{0.3\linewidth}
        \centering
        \includegraphics[width=1.0\textwidth]{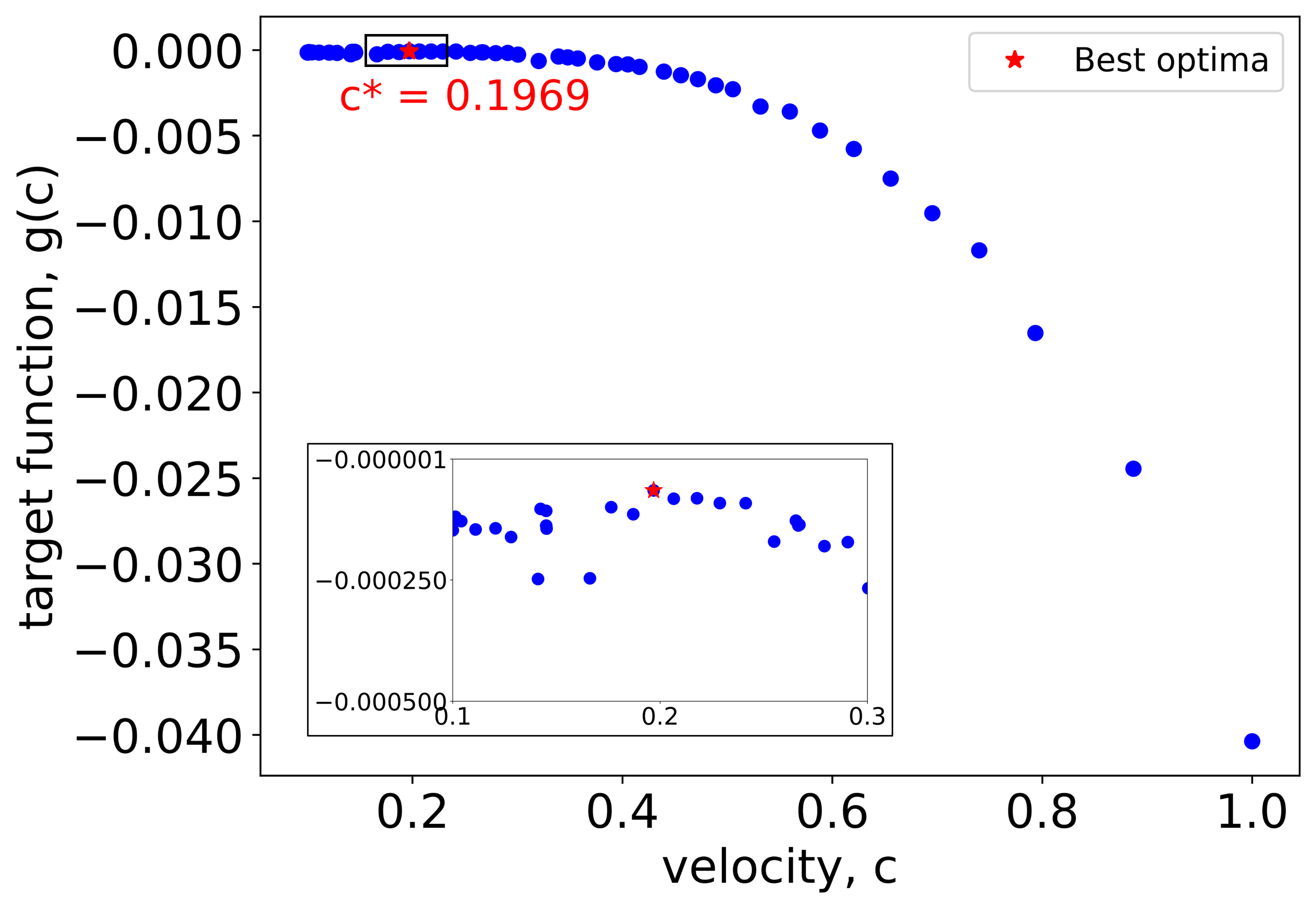}
    \end{minipage}
    \begin{minipage}[b]{0.3\linewidth}
        \centering
        \includegraphics[width=1.0\textwidth]{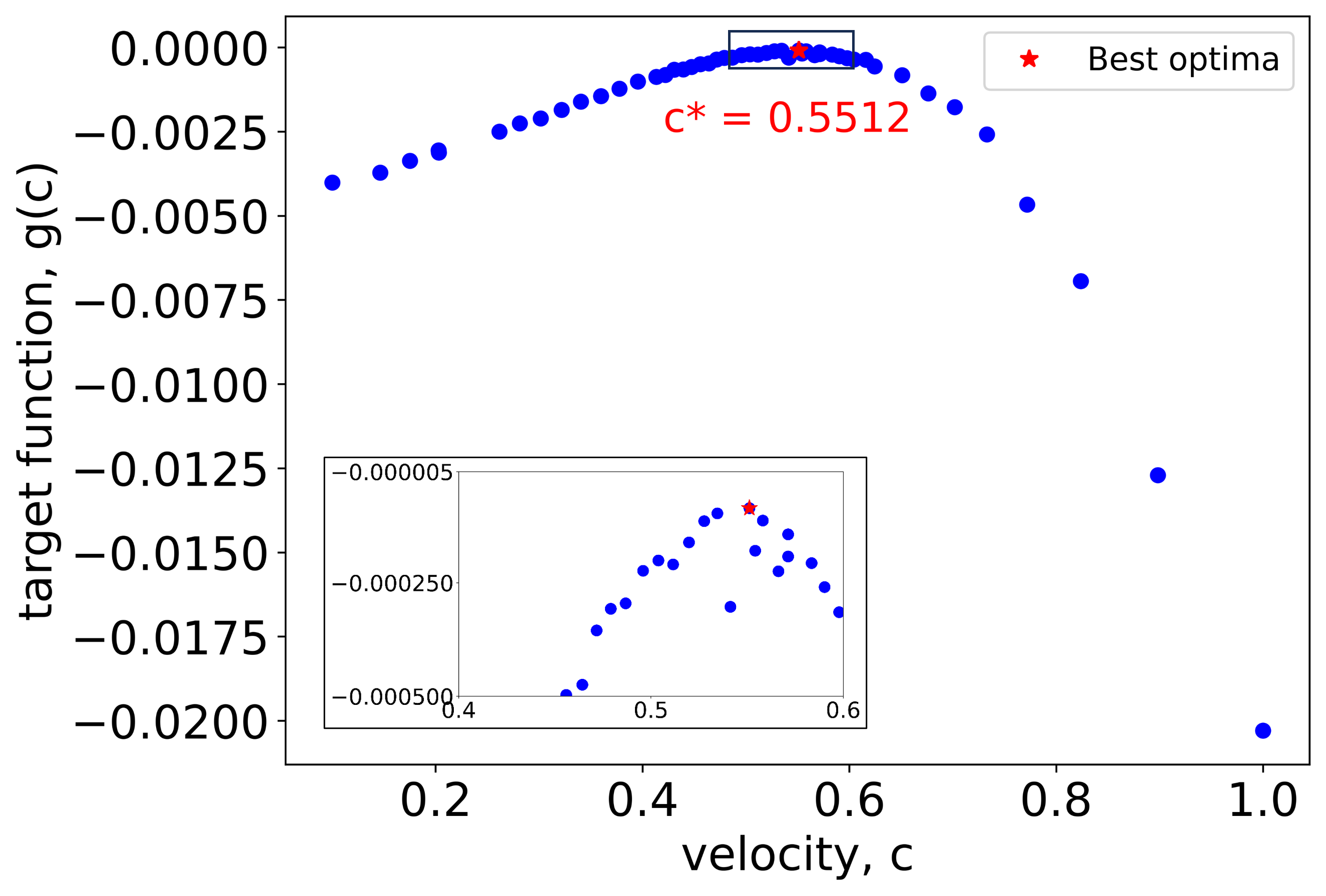}
    \end{minipage}
    \begin{minipage}[b]{0.3\linewidth}
        \centering
        \includegraphics[width=0.95\textwidth]{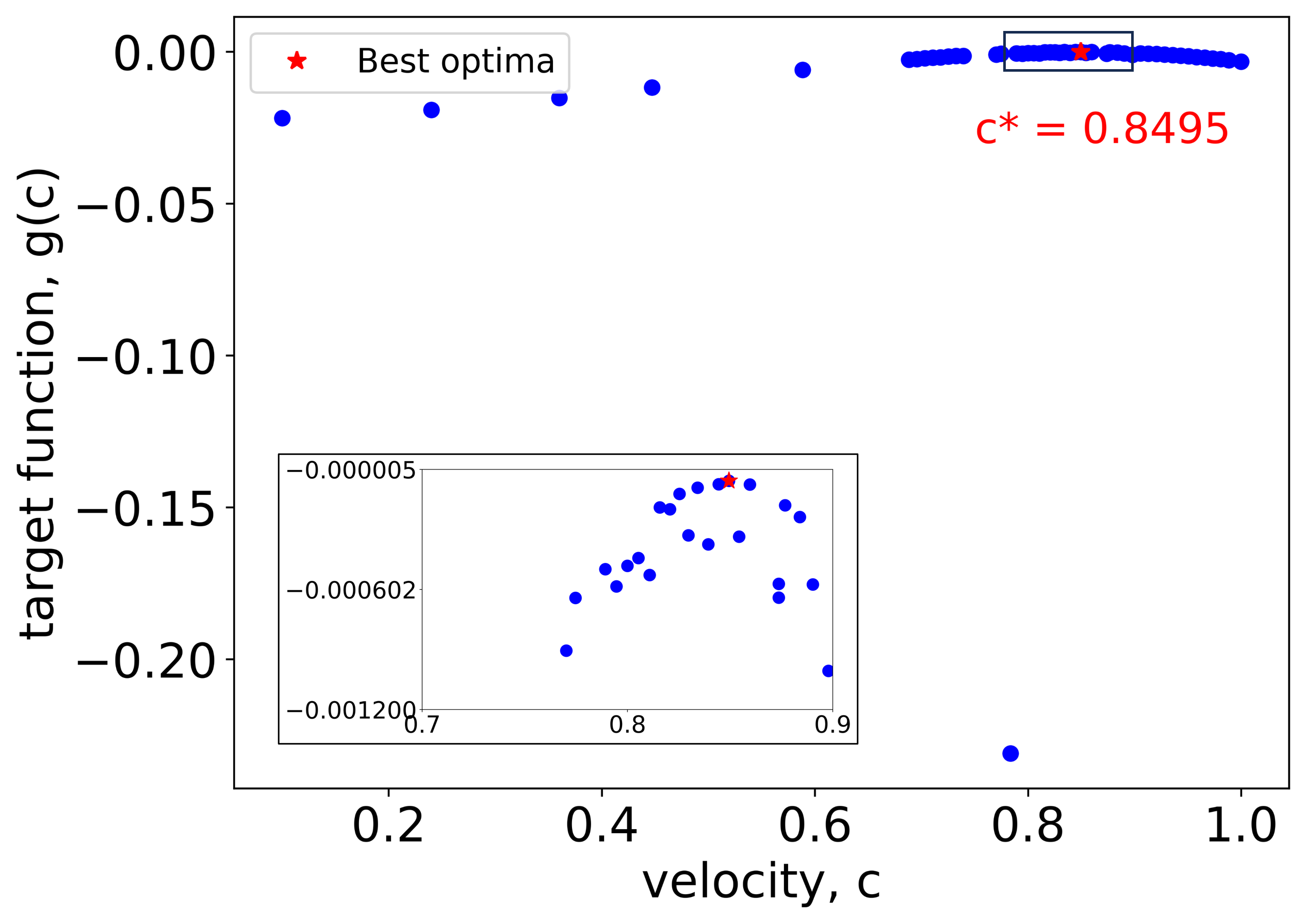}
    \end{minipage}
    \caption{Target function-wave velocity results from BOPINN for the best optimal run for all three cases.}
    \label{fig:BOPINNbestcurves}
\end{figure*}

\section{Results and Discussions} \label{sec:results}
The forward solution is set up using a PINN, represented mathematically in Eq.~\ref{eq:PINNopt}. Limited-memory Broyden–Fletcher–Goldfarb–Shanno algorithm (or L-BFGS) is used as a full-batch gradient-based optimization algorithm \cite{liu1989limited}. It is a quasi-Newton, second-order optimization method that approximates the Hessian to account for the curvature of the objective function. We have created the dataset and collocation points randomly with 25,000 points in $x \in [0,L], \hspace{2mm} t \in [0,T]$, where L = 10 m and T = 1 second. The test points are 5000. A six-layered deep neural network is utilized with 64,128,128,128,128,64 neurons with tanh activation function. In order to incorporate uncertainty in the forward PINN model, we have introduced dropouts of 10\% after every layer. The network architecture is selected based on the tradeoff between training time and the loss \cite{rautela2021simulation}. A smaller network provides faster training and helps in the inverse estimation of wave velocities under Bayesian optimization. The wave velocity in the waveguide is defined based on expert knowledge of the velocities in most of the materials as $c \in C = [1000 \hspace{1mm} 10000]$ m/s. One important point to highlight here is that we have scaled the PDE, i.e., the objective function of PINN (Eq.~\ref{eq:PINNopt}) to bring $C \in [0.1\hspace{1mm}1]$. It transforms the inputs $(x,t,c)$ of PINN in similar ranges, which helps in stabilizing L-BFGS.

The simulated data (snapshot observation at $\tilde{t}$ = 0.25 second) is obtained with an analytical model for \Cref{eq:wave1d,eq:domain,eq:ic,eq:bc} at different wave velocities i.e., 2000, 5500, 8500 m/s, i.e., 0.2, 0.55, 0.85. Each data is used for different case studies to test the parameter estimation ability of BOPINN. In order to make the simulated data more realistic, white noise of SNR = 36.34 dB is added to the data. Fig.~\ref{fig:pinnsol} shows the comparison of the PINN solution and analytical solution for c = 0.85 ($c=8500$ m/s). It can be seen that the PINN-based forward solver shows promising results and matches well with the analytical solution. 

The Bayesian optimizer sequentially queries the PINN model at different wave velocities in $C$. The target function is the mean squared error between prediction from PINN and a snapshot observation at $\tilde{t} = 0.25 s$. BO is performed for 50 iterations, out of which the first five iterations are used to formulate a prior for the Gaussian process model. For every iteration of BO, PINN solves the PDE at a particular wave velocity suggested by the acquisition function. The target function is constrained with $C \in [0.1 \hspace{1mm} 1]$, based on the realistic velocities in isotropic materials. The Upper Confidence Bound (UCB) is used as an acquisition function. The value of $\kappa$ is selected as 2.45 by trial-and-error. BO is repeated for ten different runs to capture the uncertainty in estimation.

BOPINN is tested for all three cases, i.e., three different isotropic media with true wave velocities as c = 0.2, 0.55, 0.85 are tabulated in Table~\ref{tab:BOPINNtab}. In the table, the second column represents the true wave velocity, and the rest of the columns present the optimal solution ($g(c)^*$) and optima ($c^*$) for BO across ten different runs. The best optimal solution is the maximum target function values of 10 optimal solutions obtained for 10 different runs, respectively. The mean and standard deviation across different runs are also highlighted in the table. It can be seen that the wave velocity estimates are very close to the true value. The standard deviation is nearly ten times lower than the mean, which shows that the obtained results are consistent and reliable. We have also shown the target function-wave velocity plot in Fig.~\ref{fig:BOPINNbestcurves} for the best optimal solution for all three cases. Different points of the curve represent target function and velocity values across 50 different iterations. The best optima is highlighted in red in the figures as well as in the table.

In this study, we have obtained estimation accuracies of (98.45\%, 99.78\%, and 99.95\%) and (99.2\%, 94.36\%, and 99.76\%) for the best and mean optimal results for all three cases, respectively. The BOPINN is performed for 50 iterations, where at every iteration, a PINN is solved at new $c$ suggested by BO. We have noticed an average computational time of $\sim$4 minutes per iteration, the majority of which is utilized in solving the PINN. On the other hand, the grid search algorithm requires 90 iterations for a precision of 2 significant decimal digits and 900 for a precision of 3 significant decimals. This computational advantage of BO is more evident as the number of parameters to estimate increases. One of the reasons for this advantage of BO over grid search is its adaptive search scheme. BO utilizes a surrogate Gaussian process regression to model the target function and an acquisition function to select new query points. EO algorithms also offer different adaptive search methodologies and can be compared against BO, which is one of our future research directions. In this work, we have applied BOPINN for the wave propagation problem. However, it is a more general formulation and can be used for estimating multiple parameters and fields in different types of PDEs, which is our other future research direction. 

The source code, along with the dataset will be made publicly available at the weblink*. 

\thispagestyle{github}
\section{Conclusion} \label{sec:conclusion}
In this paper, we have presented BOPINN for estimating wave propagation in isotropic materials. The method is designed to utilize the PDE learning capability of PINN and robust inverse parameter estimation of BO. We have seen that the PINN is able to learn the wave propagation behavior, which matches well with the analytical results. We have tested the capabilities of BOPINN to estimate wave velocity in three different homogeneous and isotropic media with noisy single snapshot observation. We have observed that the estimated velocity comes very close to the true value for all the test cases. The estimation accuracies are 98.45\%, 99.78\%, and 99.95\% for the three cases, respectively. The method provides the estimation uncertainty in the form of standard deviation in predicting optimal value across different runs. The algorithm can robustly and accurately estimate wave velocity in limited iterations. The proposed technique is computationally inexpensive as compared to its counterparts. The BOPINN framework decouples forward and inverse models and bypasses the retraining of the PINN model for new observations. It helps in providing a flexible architecture to perform online inversion.


\bibliographystyle{IEEEtranN}
\bibliography{mybibfile}

\end{document}